\documentclass[a4paper,11pt]{article}
\usepackage{pos}
\newcommand{\denselist}{\setlength{\itemsep}{0cm} \setlength{\parskip}{0cm}}
\usepackage{graphicx}
\usepackage{xcolor}
\title{Towards observations of nuclearites in Mini-EUSO}
\ShortTitle{Towards observations of nuclearites in Mini-EUSO}

\author*[a]{L W Piotrowski}

\author[b,c]{D Barghini}
\author[b,c]{M Battisti}
\author[e]{A Belov}
\author[b,c]{M Bertaina}
\author[b,c]{F Bisconti}
\author[f]{C Blaksley}
\author[h]{K Bolmgren}
\author[p]{F Cafagna}
\author[b,g]{G Cambi\`e}
\author[n]{F Capel}

\author[o,f,g]{M Casolino}

\author[f]{T Ebisuzaki}
\author[b,c]{F Fenu}
\author[j]{A Franceschi} 
\author[h]{C Fuglesang} 
\author[b,c]{A Golzio}
\author[f]{P Gorodetzki} 
\author[m]{F Kajino} 
\author[f]{H Kasuga}
\author[e]{P Klimov}
\author[r]{V.~Kungel} 
\author[b,c]{M Manfrin}
\author[o]{L Marcelli}
\author[k]{W Marsza{\l}}
\author[b,c]{H Miyamoto}
\author[b,c]{M Mignone} 
\author[j]{T Napolitano}
\author[q]{G Osteria}
\author[l]{E Parizot} 
\author[o,g]{P Picozza} 

\author[b]{Z Plebaniak} 
\author[l]{G Pr\'ev\^ot} 
\author[o,g]{E Reali}
\author[j]{M Ricci} 
\author[f]{N Sakaki} 
\author[k]{K Shinozaki} 
\author[k]{J Szabelski} 
\author[f]{Y Takizawa} 
\author[f]{S Wada}
\author[r]{L.~Wiencke} 

\affiliation[a]{ Faculty of Physics, University of Warsaw - Warsaw, Poland}

\affiliation[b]{ INFN, Sezione di Torino - Torino, Italy}
\affiliation[c]{ Dipartimento di Fisica, Universit\'a di Torino, Italy}
\affiliation[d]{ Faculty of Physics, M.V. Lomonosov Moscow State University - Moscow, Russia}
\affiliation[e]{ Skobeltsyn Institute of Nuclear Physics, Lomonosov Moscow State Univ. - Moscow, Russia}
\affiliation[f]{ RIKEN - Wako, Japan}
\affiliation[g]{ Universit\'a degli Studi di Roma Tor Vergata - Dipartimento di Fisica, Roma, Italy}
\affiliation[h]{ KTH Royal Institute of Technology - Stockholm, Sweden}
\affiliation[i]{ Universit\'e de Paris, CNRS, Astroparticule et Cosmologie, F-75006 Paris, France}
\affiliation[j]{ INFN-LNF - Frascati, Italy}
\affiliation[k]{ National Centre for Nuclear Research - Lodz, Poland}
\affiliation[l]{ APC, Univ Paris Diderot, CNRS/IN2P3, CEA/Irfu, Obs de Paris, Sorbonne Paris Cit\'e, France}
\affiliation[m]{ Konan University, Kobe, Japan}
\affiliation[n]{ Technical University of Munich- Munich, Germany} 
\affiliation[o]{ INFN, Sezione di Roma Tor Vergata - Roma, Italy}
\affiliation[p]{ INFN, Sezione di Bari - Bari, Italy}
\affiliation[q]{ INFN, Sezione di Napoli - Napoli, Italy}
\affiliation[r]{ Colorado School of Mines, Golden, CO, USA}

\forColl{JEM-EUSO} % W/O "Collaboration"

\emailAdd{publ@lwp.email}

\abstract{
Mini-EUSO is a small orbital telescope with a field of view of $44^{\circ}\times 44^{\circ}$, observing the night-time Earth mostly in 320-420 nm band. Its time resolution spanning from microseconds (triggered) to milliseconds (untriggered) and more than $300\times 300$ km of the ground covered, already allowed it to register thousands of meteors. Such detections make the telescope a suitable tool in the search for hypothetical heavy compact objects, which would leave trails of light in the atmosphere due to their high density and speed. The most prominent example are the nuclearites -- hypothetical lumps of strange quark matter that could be stabler and denser than the nuclear matter. In this paper, we show potential limits on the flux of nuclearites after collecting 42 hours of observations data.
}

\FullConference{37$^{\rm{th}}$ International Cosmic Ray Conference (ICRC 2021)\\
	July 12th -- 23rd, 2021\\
	Online -- Berlin, Germany}

\begin{document}
	\maketitle

	\section{Introduction}
	\label{sec:Introduction}
	
	The Universe contains objects with a wide spectrum of masses, from microscopic to the heaviest stars, which are formed from the ordinary constituents of matter such as electrons, protons and neutrons. There are, however, parallel mass spectra of more exotic objects, including heavy compact objects. One example are black holes, of which we have detected only those with stellar masses or heavier, while lighter are theorised. Another -- hypothetical -- example are so-called nuclearites.
	
	Nuclearites, which are, in principle, heavy ``strangelets'', would be formed from ``strange quark matter'' -- stable lumps of big amount of up, down and strange quarks. The addition of the strange quark, according to Witten \cite{bib:witten}, could make them more stable and denser than the ordinary matter formed from protons and neutrons. In such case, they could cross the Earth's atmosphere almost unaffected, and those with speeds of the order of 100 km/s or faster would emit light in a mechanism similar to meteors described in \cite{bib:drg}, on which the presented estimation is based. Their brightness would depend mainly on their speed, diameter and density of the atmosphere (or any other transparent medium that they would cross). The same light emission principle could be applied to other heavy compact objects if they are stable enough to survive the passage through the medium. Other mechanisms of light emission explained in \cite{bib:ss} and \cite{bib:luis} are possible and could result in the luminosity of the object lower by many orders of magnitude.
	
	Unlike meteors, nuclearites would lose almost no energy in the atmosphere and leave trails with a very small variation of brightness. If they were to be a part of Dark Matter, their average speed would be of the order of 200 km/s, much faster than the maximal speed of meteors -- 72 km/s. If they were of cataclysmic origin, their speed would likely be even higher. In addition, mainly due to their diameter, nuclearites not exceeding the mass of hundreds of kilograms would emit light much closer to the ground. Therefore, due to similarities, experiments able to observe meteors should also be able to detect nuclearites and other heavy compact objects crossing the atmosphere. Due to observational differences, they should be able to separate the populations. Orbital telescopes, such as Mini-EUSO are well suited for the detection attempts due to the sheer volume of the atmosphere observable from space.
	
	\section{Mini-EUSO}
	
	Mini-EUSO \cite{bib:Mini-EUSO} is a small orbital telescope, designed within the JEM-EUSO programme \cite{bib:JEMEUSO}, observing the night-time Earth from the International Space Station (ISS) through a UV-transparent window inside the Zvezda module. It is composed of two 25 cm diameter Fresnel lenses focusing light on a Photon Detection Module (PDM) consisting of 36 multi-anode photomultipliers (MAPMTs), encompassing 2304 pixels. The field of view is $44^{\circ}\times 44^{\circ}$, with a single-pixel side covering roughly 6 km on the ground, and the whole PDM more than 300 km. The spectral acceptance spans between 320 and 420 nm, making Mini-EUSO mostly a UV telescope. The PDM data is gathered in 3 time resolutions. D1 data consists of packets of 128 frames, each with 2.5 $\mathrm{\mu s}$ exposure, stored upon receiving a fast-events trigger from the FPGA. D2 data packet consists of 128 frames, each being an average of 128 D1 frames, forming a 320 $\mathrm{\mu s}$ block. It is collected after receiving a separate, slow-events trigger. D3 data are untriggered, forming a continuous ``movie'' with a single frame being an average of $128\times 128$ D1 frames, spanning 40.96 ms.
	
	The PDM is a very sensitive instrument and thus can be damaged by excessive light. Thus two main levels of protection were introduced. The first one switches a part (an ``EC-unit'' composed of 4 MAPMTs) of the detector to lower gain if a few very bright pixels are detected in it. This happens quite often when going over the cities, etc. The second one is an analogue over-current protection, sensitive to the summed signal in all the EC-unit pixels.
	
	The telescope is also equipped with a small near infra-red camera and a visible light camera set to take photos with 5 s exposure time, photodiodes for detecting the night/day transitions, and a small silicon photomultiplier.
	
	\section{D3 offline trigger}
	
	The D3 data frame of 40.96 ms length allows Mini-EUSO to register whole tracks of meteors, at the same time revealing the variations in brightness for the longer ones. As mentioned in sec. \ref{sec:Introduction}, nuclearites crossing the atmosphere should, up to a point, exhibit similar observational properties as meteors, thus a search for the latter is a good starting point in the search for the former.
	
	As D3 data are untriggered, a dedicated off-line trigger has been created to detect events, currently with the main purpose of finding meteors. The main steps of the trigger are the following:
	
	\begin{enumerate}
		\denselist	
		\item Divide the data into the chunks with the same gain (i. e. with and without excessive light protection)
		\item Estimate background for each pixel\footnote{The background estimation currently uses CERN ROOT \cite{bib:ROOT} TSpectrum class, but a more suitable estimator is intended to be used in the future.}
		\item Find frames over the background-based threshold for each pixel
		\item Remove (pixel,frame) pairs that do not have another such pair in a 4 frames vicinity
		\item Group the (pixel,frame) pairs across space (PDM) and time into events using a KD-tree
		\item Perform initial categorisation of events (static, slow, extended, meteor candidates, etc.)
		\item Perform additional quality cuts on meteor candidates
	\end{enumerate}
	
\begin{figure}[t]
		\begin{center}
			\includegraphics[width=0.4\textwidth]{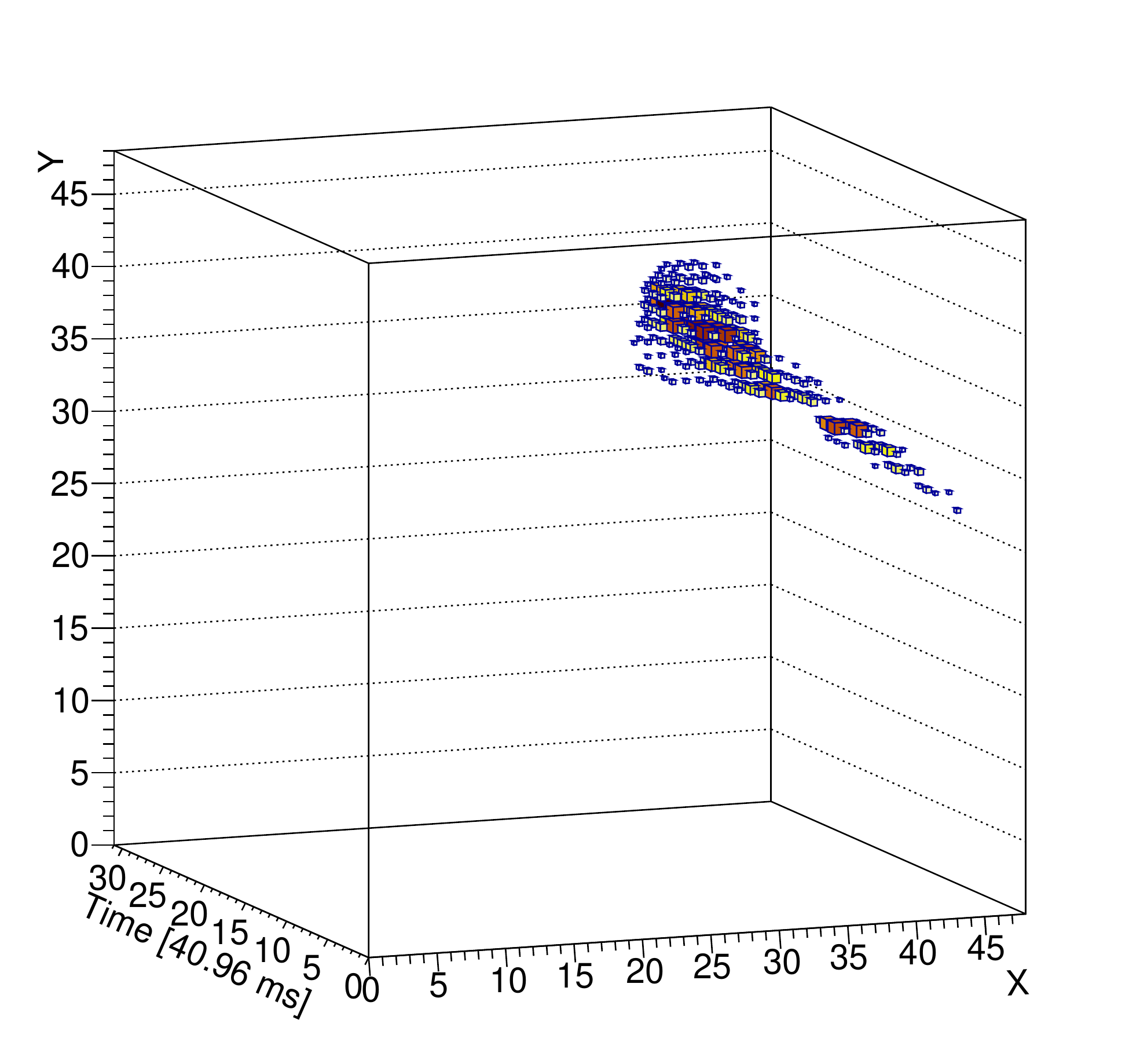}
			\includegraphics[width=0.59\textwidth]{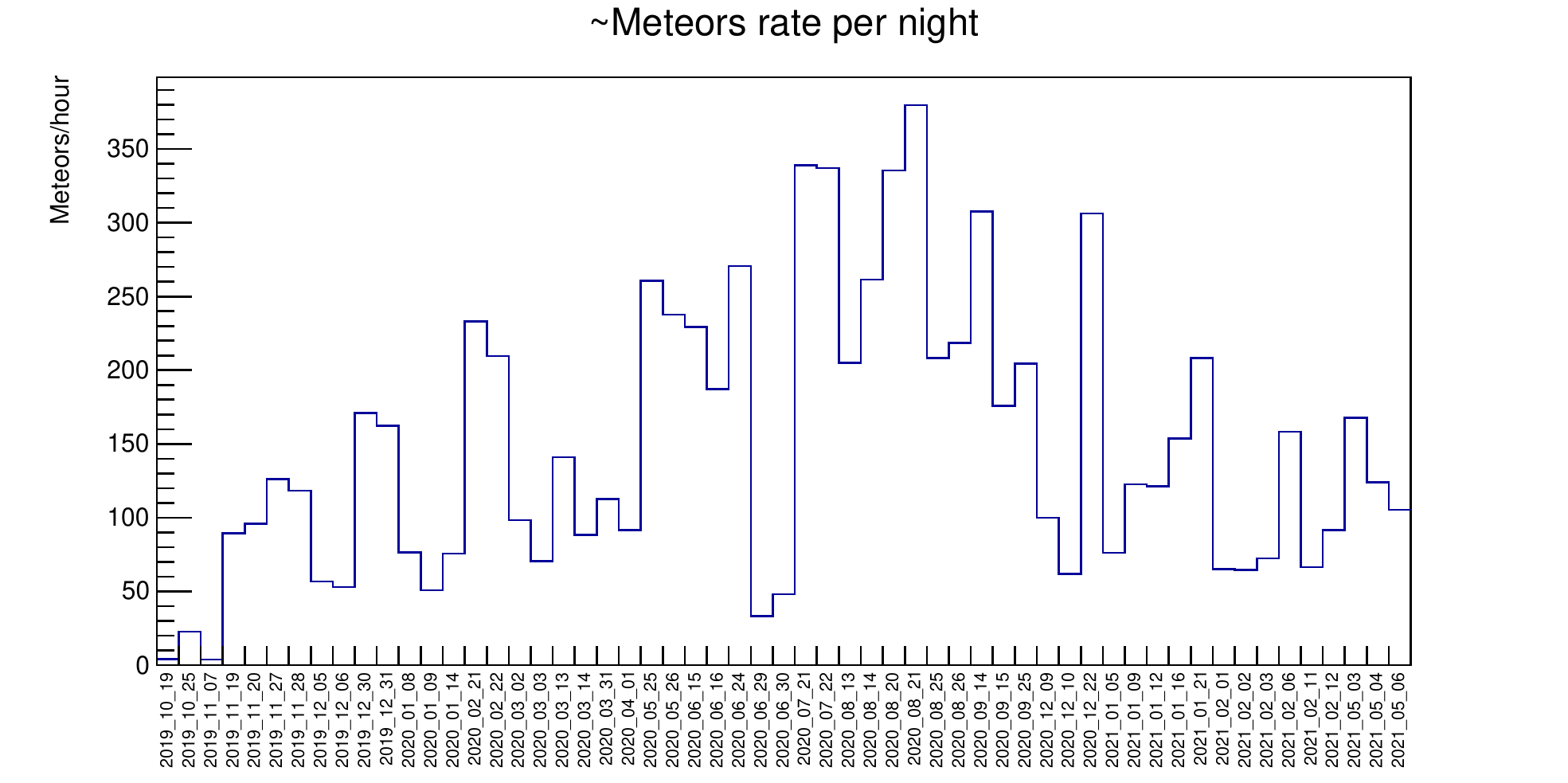}
			\caption{{\em Left:} An example of a meteor found in the Mini-EUSO data drawn in x,y,time coordinates, after background subtraction. The size of the boxes and their colour denote the number of counts deposited in each pixel at each frame. The visible break in the track is caused by the meteor crossing the dead space between MAPMTs. {\em Right:} Meteors detected by the off-line algorithm rate per hour for each Mini-EUSO observation night. The bins are not equidistant in time. The long trend is mainly due to higher meteors activity during the summer time, while shorter time-scale variations are mainly due to varying Moon phase.}
			\label{fig:meteor_example}
		\end{center}
	\end{figure}

	After these steps, basic meteor properties such as lightcurve, track on the PDM (see fig. \ref{fig:meteor_example}, left) and speed are measured with simple and quick methods and stored in a database. The quality of these measurements is far from perfect, especially for very dim meteors, where shortcomings of the current background subtraction method have the largest influence. However, the estimation serves well the purpose of showing general properties of the dataset and identifying interesting events among the bright candidates.
	
	\section{Meteor statistics}
	
%	\begin{figure}[hbt]
%		\begin{center}
%			\includegraphics[width=0.7\textwidth]{images/meteors_per_night_hour.pdf}
%			\caption{Meteors detected by the off-line algorithm rate per hour for each Mini-EUSO observation night. The bins are not equidistant in time. The long trend is mainly due to higher meteors activity during the summer time, while shorter time-scale variations are mainly due to varying Moon phase.}
%			\label{fig:meteor_rate}
%		\end{center}
%	\end{figure}

	After every data taking session, only roughly 20\% of the data is downlinked to the ground due to the ISS bandwidth limitations. The remaining data is periodically transported to Earth on USB pendrives. So far we have obtained about 63 hours of data on the ground, 42 hours without gain reduction due to over-light protection\footnote{This is a full-PDM-equivalent, as parts of the PDM -- the EC-units -- go into the protection mode independently.} -- the nominal observation mode. Almost all out of 5552 meteor candidates were identified in this part of the data. The visual inspection uncovered 616 false candidates in this set, forming the average false event rate of about 11\%. This rate is by no means constant, with the first sessions after launch containing up to 20\% background. The average rate of meteors seen by Mini-EUSO is about 2 meteors per minute (see fig. \ref{fig:meteor_example}, right) and strongly depends on the background light level varying with the observed terrain albedo, presence of ground sources and the Moon phase.
	
	None of the analysed events has a semi-constant lightcurve, long-duration and measured speed significantly over 72 km/s, which excludes obvious heavy compact objects candidates. Lack of fast events also excludes meteors of interstellar origin. The two meteors of the longest duration that has been identified are visible for 61 and 90 D3 frames, corresponding to 2.5 and 3.7 s -- long, but definitely not unheard of. It has to be noticed, however, that the longer of the two exited the field of view before becoming invisible.
	
	\section{Towards observations of nuclearites}
	
	The amount of meteors observed up to now with Mini-EUSO makes it a promising tool for the search for traces left by nuclearites and other heavy compact objects in the atmosphere. No obvious candidate has been found until now, but to claim lack of detection we would need to check if our trigger can detect semi-constant brightness, faster-than-meteors tracks in the data, and estimate its efficiency in detecting such events depending on the brightness of the object and its track orientation in the atmosphere. This is best done simulating expected track over real data, similarly to what has been done in \cite{bib:pi_nuclearites}, as nuclearites are still hypothetical. We have not performed such efficiency estimation yet and we can only very roughly estimate our sensitivity based on the detector parameters and observation time.
	
	For this estimation, we follow nuclearites parameters as described in \cite{bib:drg}, namely the magnitudes, speeds and isotropic distribution of tracks, and consider only downwards going objects. For now, the observation time is assumed to be 42 hours, although in principle heavy and thus bright nuclearites could be visible in the remaining time with lower gain. A nuclearite is approximated to emit light only below a specific altitude $h_{max}$, which increases with its mass. For masses between 1 g and 100 kg this altitude changes from 31 to 62 km, respectively. Thus the exposure estimation has to be performed separately for each mass, and consists of the following steps:
	
	\begin{enumerate}
		\denselist
		\item Calculate the surface of the field of view pyramid $S$, trimmed at the top by the $h_{max}$. The bottom of the pyramid is excluded. This is an approximation of the detector's surface for nuclearite detection.
		\item Calculate the magnitude of a nuclearite at $h_{max}/2$ (an approximation of an average nuclearite altitude) visible at the ISS orbit (400 km).
		\item Recalculate the magnitude to maximal Mini-EUSO PDM counts, assuming detection efficiency of 10\% and 60\% of light focused in one pixel, and that nuclearite crosses a pixel on average in 1/2 of D3 frame exposure time.
		\item Calculate the maximum background for which such a nuclearite-illuminated pixel would be at $3\sigma$ level (assuming a Poissonian distribution of counts in pixels).
		\item Find the fraction of time in which a median of pixels' brightness in each flat-fielded MAPMT is below this background level.
		\item Average the above value over all the MAPMTs and divide by the whole observation time to get the fraction of time in which nuclearite could be detected, $t_d$.
		\item Calculate the exposure for detection, as $E(m)=S(m)*t_d(m)*\pi$, where $m$ is the mass of a nuclearite.
	\end{enumerate}
	
\begin{figure}[hbt]
		\begin{center}
			\includegraphics[width=0.5\textwidth]{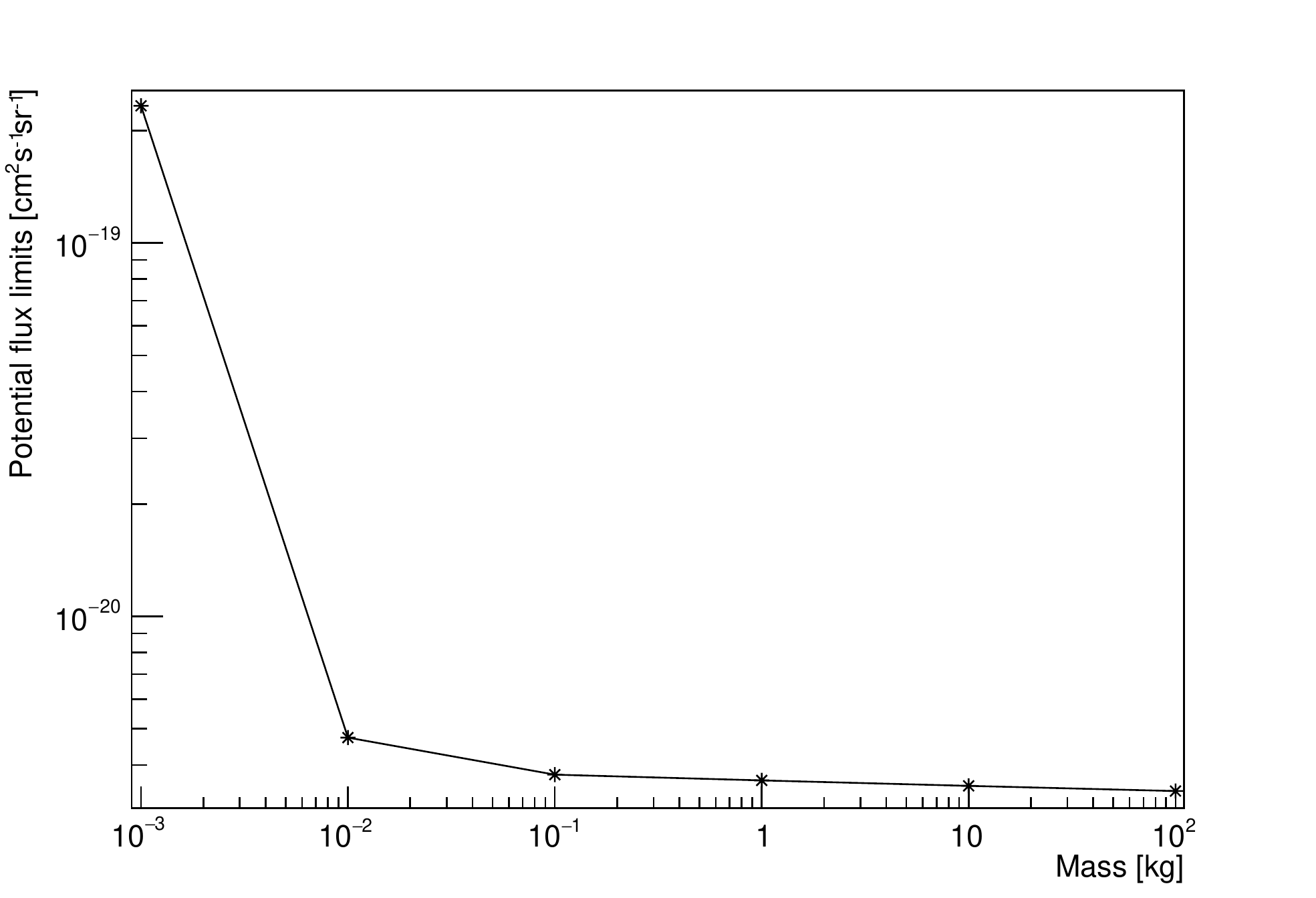}
			\caption{{\em Potential} limits for downward going nuclearite isotropic flux in the atmosphere after 42 hours of observations equivalent (see text).}
			\label{fig:flux_limits}
		\end{center}
	\end{figure}

	Assuming lack of detection of nuclearites in our data, such an exposure can be translated to a potential 90\% C. L. flux limit $\Phi(m)=2.3/E(m)$. The expected signal in pixel, the exposure and the potential flux limit are listed for each mass in table \ref{tab:nuclearite_parameters}. The potential flux limits are also shown in fig. \ref{fig:flux_limits}. The sudden change in the exposure and thus the limit between 1 g and 10 g is due to the average background in Mini-EUSO pixels. For 10 g nuclearites are already bright enough to be visible in Mini-EUSO in most conditions in the nominal observation mode, while for 1 g the background has to be very low, which happens rarely.
	
	\begin{table}
		\begin{tabular}{cccc}
			\bf Mass [kg] & Signal [pix. cnts / 2.5 $\mathrm{\mu s}$] & Exposure [$\mathrm{km^2h\ sr}$] & Pot. flux limit [$\mathrm{cm^{-2}s^{-1} sr^{-1}}$] \\
			\hline
			0.001 & 0.01 & 274272 & 2.33e-19 \\
			0.01 & 0.06 & 13495040 & 4.73e-21 \\
			0.1 & 0.3 & 16959096 & 3.77e-21 \\
			1 & 1.41 & 17571339 & 3.64e-21 \\
			10 & 6.69 & 18172560 & 3.52e-21 \\
			100 & 31.7 & 18763028 & 3.41e-21 \\
			\hline
			
		\end{tabular}
		\caption{Parameters for the potential nuclearite flux limit estimation after 42 hours of observations equivalent (see text).}
		\label{tab:nuclearite_parameters}	
	\end{table} 
	
	It has to be stressed that the {\em potential} flux limit by no means should be treated as the results of the Mini-EUSO experiment. It is based on the mere assumption of no-detection, not a real search for nuclearites in the data. Such a search has yet to be performed. Moreover, already the meteor-optimised search algorithm is much more complicated than a simple requirement of $3 \sigma$ signal-to-noise ratio in one pixel. With sufficient computing power, significant tracks consisting of non-significant pixels can be detected (so-called track-before-detect methods, such as unthresholded Hough transform). This could increase detection efficiency compared to the above estimation. On the other hand, tracks have to be of sufficient length to be considered a track, and the apparent speed of the object has to be sufficiently large, or the lightcurve has to be sufficiently long to allow distinction from meteors. That would impose cuts on the angular acceptance, which are not included in this estimation, and reduce the exposure.
	
	Other simplifications include taking the field of view pyramid surface as the detection area, while in principle one should take the pyramid cross-section perpendicular to the incoming nuclearites directions and integrate it over $2\pi$ when considering downward going, isotropic flux of objects. In addition, we approximate the efficiency of the whole detector as 10\%. While this efficiency remains to be confirmed, in final calculations we should take into account the efficiency of each pixel separately. This is easiest done with simulating tracks over the real background. All these approximations are likely to make the presented potential flux limits somewhat too optimistic, but probably not more than by a factor of a few.
	
	\section{Summary}
	
	In this paper, we have described the basic principles of the off-line trigger for events detection on the data with 40.96 ms time resolution. The trigger, at the moment, focuses on linear tracks. It has detected 5552 events in 42 hours of data, of which 4963 has been initially confirmed as meteors after visual inspection. That gives an average rate of about 2 meteors per minute of observations in the nominal Mini-EUSO mode. This rate varies significantly with background light intensity.
	
	Based on the observation time, detector's field of view and the orbit of the ISS, we have estimated exposures for detection of nuclearites in the mass range between 1 g and 100 kg. Assuming lack of detection, we have estimated the potential limits on the flux of nuclearites. However, the real flux limits will be provided after performing a dedicated search for nuclearites and estimating the trigger efficiency for nuclearite detection.

	\begin{acknowledgments}
	
This work was partially supported by Basic Science Interdisciplinary Research Projects of 
RIKEN and JSPS KAKENHI Grant (22340063, 23340081, and 24244042), by  the Italian Ministry of Foreign Affairs	and International Cooperation, by the Italian Space Agency through the ASI INFN agreements n. 2017-8-H.0 and n. 2021-8-HH.0, by NASA award 11-APRA-0058, 16-APROBES16-0023, 17-APRA17-0066, NNX17AJ82G, NNX13AH54G, 80NSSC18K0246, 80NSSC18K0473, 80NSSC19K0626, and 80NSSC18K0464 in the USA,   by the French space agency CNES, by the Deutsches Zentrum f\"ur Luft- und Raumfahrt, the Helmholtz Alliance for Astroparticle Physics funded by the Initiative and Networking Fund of the Helmholtz Association (Germany), by Slovak Academy of Sciences MVTS JEM-EUSO, by National Science Centre in Poland grants 2017/27/B/ST9/02162 and 2020/37/B/ST9/01821, by Deutsche Forschungsgemeinschaft (DFG, German Research Foundation) under Germany's Excellence Strategy - EXC-2094-390783311, by Mexican funding agencies PAPIIT-UNAM, CONACyT and the Mexican Space Agency (AEM), as well as VEGA grant agency project 2/0132/17, and by by State Space Corporation ROSCOSMOS and the Interdisciplinary Scientific and Educational School of Moscow University "Fundamental and Applied Space Research".
	
	\end{acknowledgments}

% Full authors list (ONLY FOR COLLABORATIONS)

\clearpage

\section*{Full Authors List: JEM-EUSO\ Collaboration}

\scriptsize

\noindent
G.~Abdellaoui$^{ah}$, 
S.~Abe$^{fq}$, 
J.H.~Adams Jr.$^{pd}$, 
D.~Allard$^{cb}$, 
G.~Alonso$^{md}$, 
L.~Anchordoqui$^{pe}$,
A.~Anzalone$^{eh,ed}$, 
E.~Arnone$^{ek,el}$,
K.~Asano$^{fe}$,
R.~Attallah$^{ac}$
% ...

\vspace*{.3cm}

\noindent
% Algeria (Dezember 2013) - 7 institutes
% ...
$^{ac}$ LPR at Dept. of Physics, Faculty of Sciences, University Badji Mokhtar, Annaba, Algeria \\
% ...
$^{ah}$ Telecom Lab., Faculty of Technology, University Abou Bekr Belkaid, Tlemcen, Algeria \\
% Bulgaria ready (02042012)  - 1 institutes 
% ...
$^{cb}$ Universit\'e de Paris, CNRS, AstroParticule et Cosmologie, F-75013 Paris, France\\
% ...
$^{ed}$ Istituto Nazionale di Fisica Nucleare - Sezione di Catania, Italy\\
% ...
$^{eh}$ INAF - Istituto di Astrofisica Spaziale e Fisica Cosmica di Palermo, Italy\\
% ...
$^{ek}$ Istituto Nazionale di Fisica Nucleare - Sezione di Torino, Italy\\
$^{el}$ Dipartimento di Fisica, Universita' di Torino, Italy\\
% ...
$^{fe}$ Institute for Cosmic Ray Research, University of Tokyo, Kashiwa, Japan\\ 
% ...
$^{fq}$ Nihon University Chiyoda, Tokyo, Japan\\ 
% ...
$^{md}$ Universidad Polit\'ecnia de madrid (UPM), Madrid, Spain\\
% ...
$^{pd}$ University of Alabama in Huntsville, Huntsville, AL; USA\\
$^{pe}$ Lehman College, City University of New York (CUNY), NY, USA\\
% ...

\end{document}